\documentclass{elsart}
\usepackage{epsfig}
\usepackage{amsmath}
\usepackage{amssymb}
\usepackage{amsfonts}
\newcommand{\TB}{\ensuremath{\tan\beta}}
\newcommand{\as}{\ensuremath{\alpha_s}}
\newcommand{\Obig}[1]{\ensuremath{\mathcal{O}(#1)}}
\newcommand{\GeV}{{\rm GeV}}
\newcommand{\DR}{\ensuremath{\overline{\rm DR}}}

\begin{document}
\begin{frontmatter}
\title{On the two-loop ${\mathcal O}(\alpha_s^2)$ corrections to the
pole mass of the $t$-quark in the MSSM}
\author[bltp]{A.~Bednyakov},
\ead{bednya@thsun1.jinr.ru}
\author[bltp,itep]{D.I.~Kazakov},
\author[bltp]{A.~Sheplyakov},
\ead{varg@thsun1.jinr.ru}
\ead[url]{http://theor.jinr.ru/\~{}varg}
\address[bltp]{Joint Institute for Nuclear Research, Dubna, Russia}
\address[itep]{Institute for Theoretical and Experimental Physics, Moscow, Russia}

\begin{abstract}
	The paper is devoted to the calculation of additional two-loop $\Obig{\as^2}$
	MSSM corrections to the relation between the pole mass of the $t$-quark and
  its running mass in the $\DR$ scheme. Firstly, a contribution from axial
  part of the quark self-energy, which was erroneously omitted in our earlier
  work, is obtained. Secondly, the value of the second order
  contribution from large mass expansion in $m_t/M_{SUSY}$ is studied.
  Finally, two-loop anomalous dimension of the running quark $\DR$-mass 
	in the supersymmetric QCD is calculated.
\end{abstract}

\end{frontmatter}

\begin{section}{Introduction}
	Two-loop $\Obig{\as^2}$ MSSM corrections to the relation
	between the pole mass of the $t$-quark and its running
	$\DR$ mass had been calculated in \cite{Bednyakov:2002sf} by
	means of the large mass expansion \cite{AsympExpansion} in small parameter
	$m_t/M_{hard}$, where $M_{hard}$ stands for all mass scales involved in the
	problem that are much larger than $m_t$.

	The initial idea of this paper was to provide more
	terms in this expansion and to study 
	the influence of these terms on the final result.
	We restricted ourselves to the terms   
	$\Obig{m_t^2/M_{hard}^2}$. 
	Contrary to our initial expectations it was found that
	these terms are {\it negligible} (they affect
	the result of \cite{Bednyakov:2002sf} approximately by 0.1\%).

	In our previous calculations \cite{Bednyakov:2004gr} 
	we have reproduced the result obtained 
	in \cite{Bednyakov:2002sf} and it was found that 
	the non-zero axial contribution to quark self-energy 
	had not been taken into account by the authors of
	\cite{Bednyakov:2002sf}. This inconsistency has been
	fixed in this paper, but as numerical analysis showed 
	axial contribution is also negligible.  
	Nevertheless, the result for the relation between
	the pole mass of the $t$-quark and its running mass in
	the $\DR$ scheme presented here is free of these errors.

	Another issue studied in this paper is the calculation
	of two-loop anomalous dimension of the running quark mass
	in the supersymmetric QCD (a subset of MSSM, relevant to
	the calculation of $\as^2$ corrections). 
	It was found that contrary to the 
	non-supersymmetric case the bare mass of a quark 
	considered in the QCD sector of the MSSM can be
	used for extraction of anomalous dimension of the quark mass,
	i.e. assuming its independence 
	on renormalization scale $\bar\mu$ and differentiating 
	it with respect to $\log \bar \mu$,  
	we acquire correct expression for the two-loop anomalous dimension.
\end{section}

\begin{section}{The pole mass of the $t$-quark}

The pole mass of a particle is defined as a real part of the complex
pole of the resumed propagator  (we discuss only perturbative effects).
The full connected propagator of a quark 
can be written as
\begin{equation}
\label{fullprop}
\frac{i}{\hat{p} - m - \Sigma(\hat{p}, m_{i})},
\end{equation}
where
\begin{equation}
\Sigma(\hat{p}, m_i) = \hat{p} \Sigma_V(p^2, m_i^2) 
	+ \hat{p} \gamma_5 \Sigma_A(p^2, m_i^2)
	+ m \Sigma_S(p^2, m_i)
\end{equation}
is the self-energy of the quark, so the pole mass $M_{pole}$
satisfies the following equation:
\begin{eqnarray}
\label{polemassdef}
\left(1+\Sigma_V(M_{pole}^2, m_i^2)\right)^2 M_{pole}^2 
	& - & \Sigma_A^2(M_{pole}^2, m_i^2) M_{pole}^2 \nonumber \\
	& - & m^2 \left(1-\Sigma_S(M_{pole}^2, m_i)\right)^2 = 0.
\end{eqnarray}
Solving this equation perturbatively, one gets
\begin{eqnarray}
\label{pole2bare}
&& \frac{M_{pole}-m}{m} = \alpha M^{(1)} + \alpha^2 M^{(2)}, \quad \mbox{where} \\
&& M^{(1)} = \Sigma_V^{(1)}(m^2, m_i^2) + \Sigma_S^{(1)}(m^2, m_i), \\
&& M^{(2)} = \Sigma_V^{(2)}(m^2, m_i^2) + \Sigma_S^{(2)}(m^2, m_i) +
\frac{1}{2} {\Sigma_A^{(1)}}^2(m^2, m_i^2) \nonumber\\
&& + M^{(1)} \left( \Sigma_V^{(1)}(m^2, m_i^2) 
	+ 2 m^2 \frac{\partial}{\partial p^2}
	\left( \Sigma_V^{(1)}(p^2, m_i^2) + \Sigma_S^{(1)}(p^2, m_i) \right)|_{p^2=m^2}\right)
\end{eqnarray}
and $\alpha$ stands for all couplings of the theory.
Using eq. \eqref{pole2bare}, one calculates the relation between pole
and running masses of the $t$-quark. In our case mass parameter $m$
corresponds to the running mass $m_t(\bar \mu)$ defined in the modified
\mbox{$\DR$} scheme \cite{Avdeev:1997sz}.	
	In the MSSM strong interactions distinguish left- and 
	right-handed particles so at the two-loop order we have to take into
	account the axial part of quark self-energy $\Sigma_A^{(1)}$. 
A quantity we want to compute is defined in the  following way
\begin{equation}
\label{pole2DR}
\frac{\Delta m_t}{m_t} \equiv \frac{M_t^{pole} - m_t(\bar \mu)}{m_t(\bar \mu)}.
\end{equation}
	The lagrangian of the supersymmetric QCD and relevant
	diagrams that contribute to the quark self-energy 
	can be found in \cite{Bednyakov:2002sf}. 
	To evaluate these diagrams the large mass expansion has been used. 
	According to its prescription 
	asymptotic expansion of a Feynman integral 
	$F_\Gamma$ which depends on the large
	masses $M_1, M_2, \ldots,$ small masses $m_1, m_2, \ldots,$ and
	small external momenta $p_1, p_2, \ldots$ can be expressed as follows
	\cite{AsympExpansion}:
\begin{equation}
\label{l:asympexp}
F_\Gamma(p_1, \ldots, M_1, \ldots, m_1, \ldots) =
\sum_\gamma F_{\Gamma/\gamma}(p, m) \mathcal{M}_\gamma(p_\gamma, m)
F_\gamma(M, m, p_\gamma),
\end{equation}
where the operator $\mathcal{M}_\gamma(p_\gamma, m)$ performs Taylor
expansion in small external (with respect to the subgraph~$\gamma$)
momenta and masses. The sum runs over all asymptotically irreducible
subgraphs of the original graph $\Gamma$. 
	The reason of such a complication is that naive expansion 
	of a Feynman integral in 
	small parameters produces spurious IR divergences which have to be 
	subtracted by adding proper counter-term diagrams.

	Using two simple facts
\begin{enumerate}
	\item There can be only even number of superparticle lines in a single
			vertex, since MSSM lagrangian is R-invariant,
	\item All superparticles are considered as heavy in our problem,
\end{enumerate}
	one can prove that only three types of subgraphs are possible in
	our problem:
\begin{enumerate}
	\item All lines of a diagram are hard (Fig.~\ref{expansion:1}).
	\item A diagram with one light line. 
	      All other lines are hard (Fig.~\ref{expansion:2}).
	\item A diagram has one cut composed of two light
		lines\footnote{expansion of such a diagram was discussed in 
		detail in \cite{Berends:1996gs}}. 
		All other lines are hard (Fig.~\ref{expansion:3}). 
\end{enumerate}

  Expression \eqref{pole2DR} can be written as a Taylor series in
	$m_t/M_{hard}$:
	\begin{eqnarray} 
		\label{defSigma}
		 \frac{\Delta m_t}{m_t} && 
		 = 1 + \sum\limits_{i=1}^\infty
		           \as^i \sum\limits_{n=-1}^{\infty} m_t^n \bar\sigma_i^{(n)}
		\nonumber \\ && 
		= 1 + \as \sum\limits_{n=-1}^{\infty} m_t^n \bar\sigma_1^{(n)}
		    + \as^2 \sum\limits_{n=-1}^{\infty} m_t^n \bar\sigma_2^{(n)}
				+ \Obig{\as^3},
	\end{eqnarray}
	where $M_{hard}$ is absorbed into $\bar\sigma_i^{(n)}$. We restrict
	ourselves to $\Obig{m_t^2/M_{hard}^2}$ terms, thus,

	\begin{equation} 
		\label{pole2DRApprox}
		 \frac{\Delta m_t}{m_t} \approx 1 
		 + \as \sum\limits_{n=-1}^{2} \sigma_1^{(n)}
		    + \as^2 \sum\limits_{n=-1}^{2} \sigma_2^{(n)}
	\end{equation}
	where $\sigma_i^{(n)} \equiv m_t^n \bar\sigma_i^{(n)}$.

	We performed the calculation of \eqref{pole2DRApprox} in a 
	semi-automatic fashion. First, FeynArts \cite{Hahn:2001rv} is used to
	generate the diagrams. Then the large mass expansion of individual diagrams
	is done by means of the C++ library {\it prop2exp} \cite{prop2exp}, based 
	on GiNaC \cite{Bauer:2000cp}. The {\it prop2exp} library performs large mass 
	expansion according to \eqref{l:asympexp}, thus, calculation is reduced
	to evaluation of 2-loop vacuum integrals and 1-loop on-shell propagator
	type integrals. This task is done by the {\it bubblesII} \cite{bubblesII}
	C++ library, which recursively reduces 2-loop vacuum integrals to 
	a master-integral \cite{Davydychev:1992mt} by integration by parts
	method \cite{IntegrByParts}.
	All renormalization constants needed for acquiring 
	finite answer can be found in \cite{Bednyakov:2002sf}.
	
\begin{figure}
\begin{center}
\mbox{\includegraphics[scale=1]{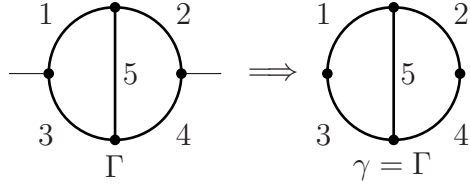}}
\caption{The diagram is composed of hard lines only; therefore,
the naive expansion can be used.}
\label{expansion:1}
\end{center}
\end{figure}

\begin{figure}
\begin{center}
\includegraphics[scale=1]{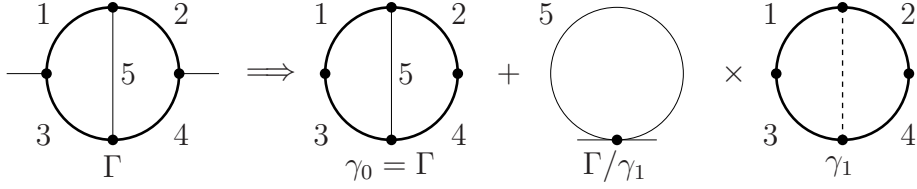}
\caption{The diagram has one light line, as a consequence,
non-trivial subgraph $\gamma_1$ is also needed. }
\label{expansion:2}
\end{center}
\end{figure}

\begin{figure}
\begin{center}
\includegraphics[scale=1]{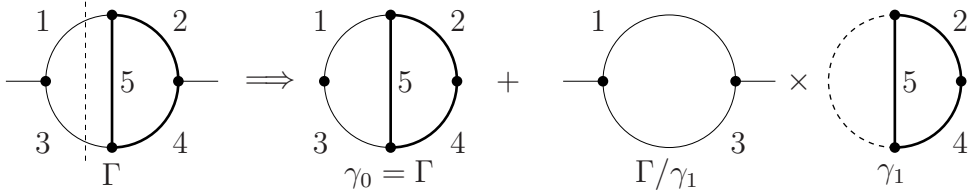}
\caption{The diagram has one cut composed of two light lines, 
so non-trivial subgraph $\gamma_1$ has to be taken
into account.}
\label{expansion:3}
\end{center}
\end{figure}
\end{section}

\begin{section}{The running mass of the $t$-quark and its anomalous dimension}
	
	As was mentioned above running \mbox{$\overline{\rm DR}$-mass}
	of a quark is defined as renormalized quark mass 
	in modified $\overline{\rm DR}$ renormalization scheme. 
	In supersymmetric QCD
	we have the following relation between bare quark mass $m_0$ and
	its running \mbox{$\overline{\rm DR}$-mass} $m(\bar \mu)$
	at the scale $\bar \mu$:
\begin{equation}
	m_0  =  m(\bar \mu) 
	\left\{
		1 + 
		\frac{\as}{4\pi}  
		\frac{\delta Z_{m}^{(1)}}{\varepsilon} 
		+ \frac{\as^2}{(4\pi)^2} 
		\left(
			\frac{\delta Z_{m}^{(2,1)}}{\varepsilon}
			+\frac{\delta Z_{m}^{(2,2)}}{\varepsilon^2}
		\right)
	\right\}, 
\end{equation}
	where $\as=g_s^2/(4 \pi)$, 
\begin{eqnarray}
\delta Z_{m}^{(1)} & = & 
		- 2 C_F, \\
\delta Z_{m}^{(2,1)} & = & 
	     \phantom{-} 6C_F - 3C_A C_F + 2C_F^2, \\
\delta Z_{m}^{(2,2)} & = & 
	    -6C_F + 3C_A C_F + 2C_F^2 
\end{eqnarray}
	and $C_F=4/3, C_A=3$ are casimirs of $SU(3)$. 
	Anomalous dimension of the quark mass $\gamma_m$ is defined as
\begin{equation}
	\frac{d}{d \log{\bar \mu^2}} m (\bar \mu) = \gamma_m m (\bar \mu)
\end{equation}
	We consider two
	ways of obtaining quark mass anomalous dimension.
	As we know physical quantities do not depend on 
	the scale parameter $\bar \mu$ so it is possible 
	to find perturbative expansion of $\gamma_m$ in
	coupling constants of the theory by differentiation 
	of pole mass expressed in terms of running parameters
	\eqref{pole2bare}. Also it is possible to extract 
	mass anomalous dimension from bare mass assuming 
	that it is renormalization group-invariant. 
	In \cite{Avdeev:1997sz} it was noticed that 
	in non-supersymmetric QCD in modified 
	$\DR$ scheme suggested by the authors 
	of that work, these two definitions 
	do not produce the same result. 
	In this paper we checked that in supersymmetric
	QCD two procedures mentioned above renders the same
	result in spite of the fact that we had used 
	the same renormalization prescription as in \cite{Avdeev:1997sz}.
	The reason of such a coincidence turns to be in the fact
	that in supersymmetric QCD gauge coupling and so-called
	evanescent coupling of $\varepsilon$-scalars renormalize 
	in the same way, so we can set them equal to each other at 
	any scale $\bar \mu$.  

	The final result for anomalous dimension of the quark mass
	in supersymmetric QCD in $\DR$ looks like
\begin{equation}
\gamma_m = 
-2 C_F \frac{\as}{4\pi}
		   +
	\left(
	12 C_F - 6 C_F C_A + 4 C_F^2 
	\right) \frac{\as^2}{ (4 \pi)^2 }
\label{anomdimen}
\end{equation}

	In a general theory with spontaneous gauge symmetry breaking
	Yukawa beta-functions do not coincide with anomalous
	dimension of corresponding fermion mass, see e.g. \cite{Jegerlehner:2003sp}.
  However, since the tree-level MSSM Higgs potential does not depend on strong
	coupling constant $\as$, and no loop corrections give contributions 
	proportional to $\as$ only (intuitively: the vacuum is colorless),
	the $\as^n$ coefficients in the perturbative expansion of \eqref{anomdimen}
	should be identical to the corresponding terms of top quark Yukawa
	coupling beta-function.
	Comparing \eqref{anomdimen} with two-loop MSSM top quark 
	Yukawa coupling beta-function \cite{Castano:1993ri,Martin:1993zk}, one can 
	see that $\as^2$ term of that beta-function indeed coincides with the 
	corresponding term of the anomalous dimension \eqref{anomdimen}. 
	This gives an additional confirmation of the correctness of our result. 
\end{section}

\begin{section}{Numerical results}
The analytical result of our calculation is complicated due
to the presence of a large number of masses, and no phenomenologically
acceptable limit seems to exist.  Therefore, we present here the
numerical analysis of our results\footnote{The full answer is avaliable
at http://theor.jinr.ru/\~{}varg/dist in a form of C++ library}.
	\begin{figure}
		\includegraphics{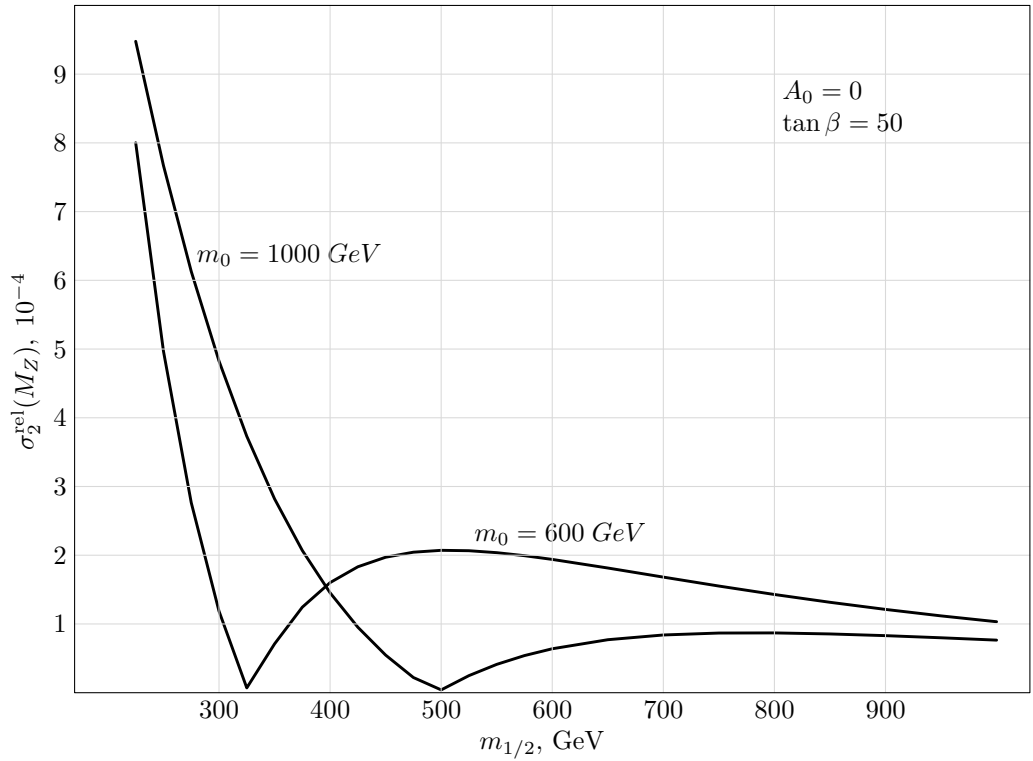}
		\caption{Relative value of second-order term of large mass expansion
		$\sigma_2^{\mathrm{rel}}$ (see \eqref{sigmaRelDef}) as a function of $m_{1/2}$.}
		\label{m12plotrel}
	\end{figure}
First, let us discuss the relative value of second order terms of large mass
expansion, i.e.
\begin{equation}
	\label{sigmaRelDef}
	\sigma_2^{\mathrm{rel}} \equiv \left| \frac{\sigma_2^{(1)}(M_Z) +
  \sigma_2^{(2)}(M_Z)}{\sigma_2^{(-1)}(M_Z) + \sigma_2^{(0)}(M_Z)} \right|.
\end{equation}
We investigated this quantity in the following regions of the CMSSM
parameter space
\begin{eqnarray*}
 && 600 \; \GeV \leq m_0  \leq 1000 \; \GeV \\
 && 200 \; \GeV \leq m_{1/2} \leq 1000 \; \GeV \\
\end{eqnarray*}
for $a_0 =  0, \; 0.4 m_0, \; -0.4 m_0$ and $\tan\beta = 33, \; 50$.
We only consider $\mu > 0$ and large values of $\tan\beta$, since small
$\tan\beta$ and negative $\mu$ seem to be excluded by experimental data
\cite{MSSMfits}. We found that $\sigma_2^{\mathrm{rel}} \lesssim 10^{-3}$ in
these regions. Typical behaviour of $\sigma_2^{\mathrm{rel}}$ is shown in
Fig.~\ref{m12plotrel}. Thus, contribution of second order large mass
expansion terms to the relation between the pole and running masses 
\eqref{pole2DR} is {\bf negligible}.

Numerical values of running SUSY parameters at the $M_Z$ scale have
been calculated as a function of CMSSM parameters with heavily modified 
version of SOFTSUSY \cite{Allanach:2001kg} in the framework of
the mSUGRA supersymmetry breaking scenario.

	\begin{figure}
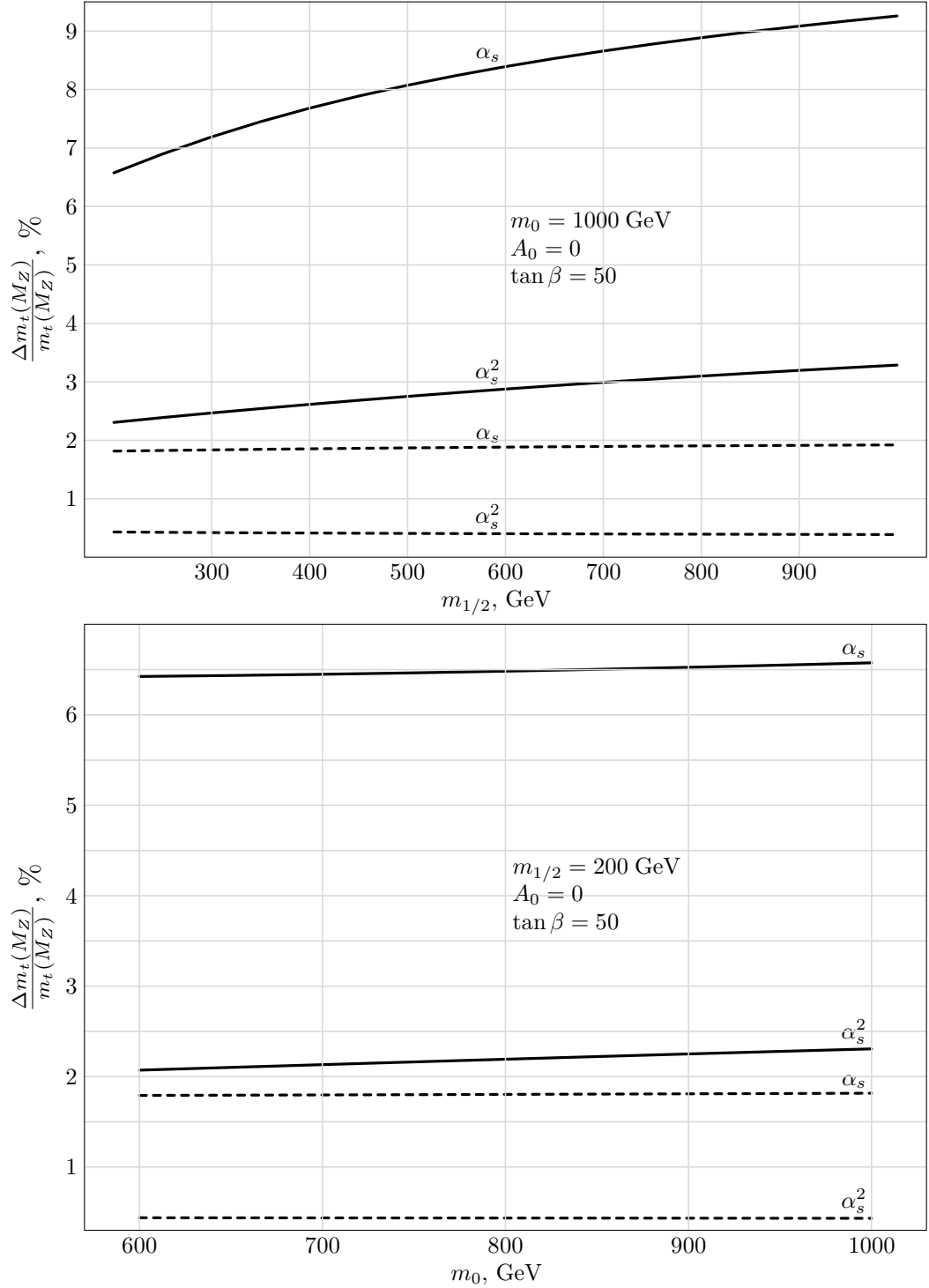

	\includegraphics{m12plotcmp}
	\includegraphics{m0plotcmp}
	\caption{ Two-loop ($\as^2$) and one-loop ($\as$) SQCD contributions to 
	$\Delta m_t(M_Z)/m_t(M_Z)$. Solid lines correspond to SQCD corrections,
	dashed line correspond to ``pupe QCD'' ones.}
	\label{m12plotcmp}
\end{figure}
In the original SOFTSUSY code (as of version 1.9) two-loop SQCD
contribution to the relation \eqref{pole2DR} is neglected, while
two-loop QCD contribution to that relation is taken into account.
This approximation is not applicable in the region of CMSSM parameters we
considered. Figure~\ref{m12plotcmp} demonstrates this.
Fig.~\ref{m12plotcmp} shows $m_{1/2}$- and $m_0$-dependence of the
$\Obig{\as^2}$ supersymmetric QCD corrections to \eqref{pole2DR}
(here and in what follows all SQCD corrections include contributions
from ``pure QCD'' diagrams). For 
comparison we plotted 1-loop SQCD contribution and 2-loop QCD contribution.
One can see that 
\begin{enumerate}
  \item two-loop SQCD correction is about of 30\% of the one-loop one,
  \item two-loop SQCD contribution is of the same order of magnitude as 
    one-loop QCD correction,
  \item two-loop QCD contribution is about of 20\% of total two-loop
		SQCD correction.
\end{enumerate}
Thus, two-loop SQCD correction to the relation between the pole and running
masses of the $t$-quark is not negligible and should be taken into account
in phenomenological analysis of MSSM\footnote{Alternatevly, if one
neglects 2-loop SQCD correction, one can neglect 2-loop QCD correction as
well}.

	\begin{figure}
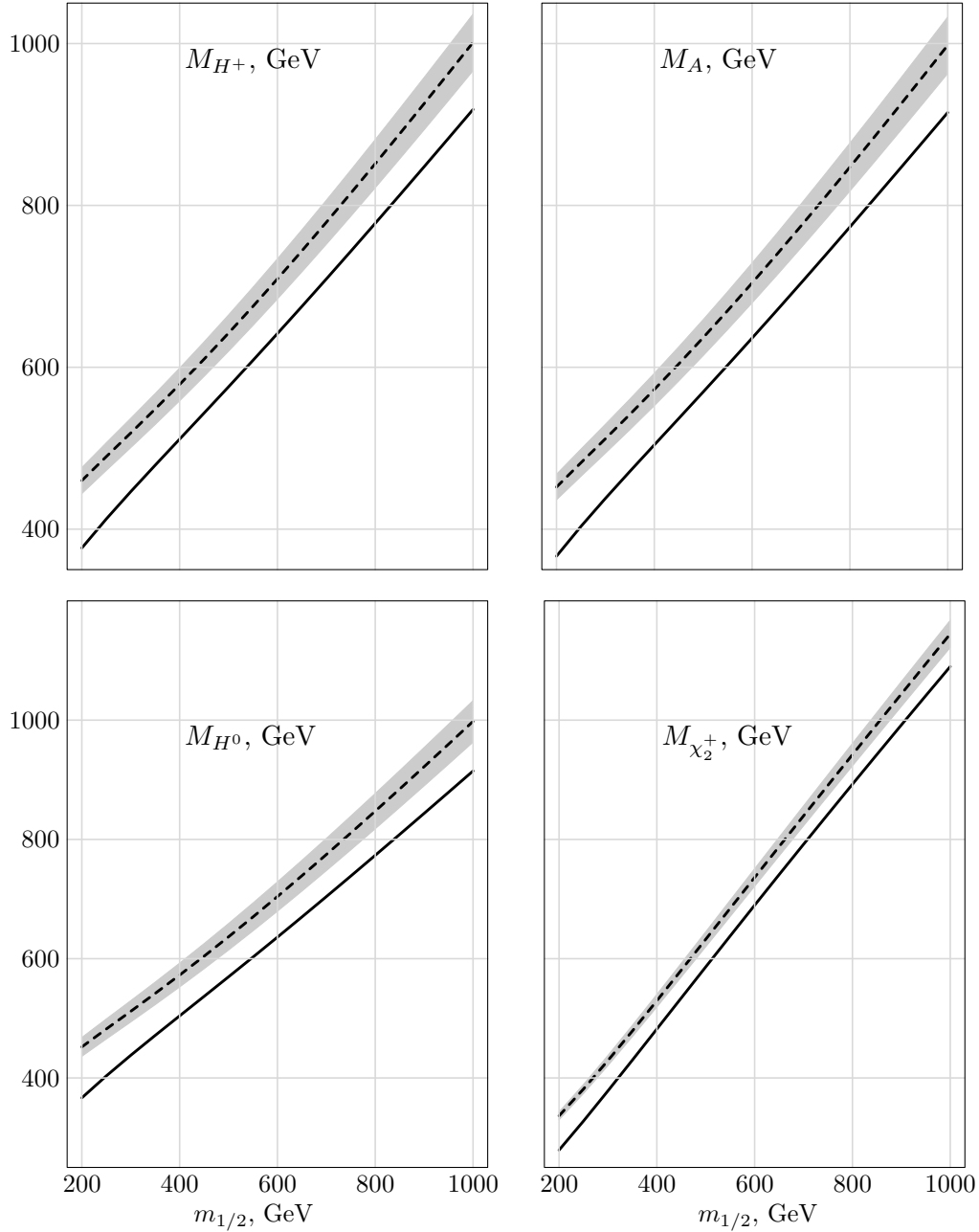

    \begin{tabular}{ll}
			\includegraphics{mHp-spectrum} & \includegraphics{mA0-spectrum} \\
			\includegraphics{mH0-spectrum} & \includegraphics{mcha2-spectrum}
   \end{tabular}
    \caption{Masses of heavy Higgs bosons and chargino as a functions 
		of $m_{1/2}$. Dashed lines correspond to the output of the original
		SOFTSUSY code
    (two-loop SQCD contribution to $\Delta m_t(M_Z)/m_t(M_Z)$ is neglected), 
    solid lines correspond to the same code {\it with} above mentioned SQCD
    corrections. Gray regions represent discrepancies between the values of
		masses predicted by different programs for calculation of MSSM mass 
		spectrum \cite{Belanger:2005jk}. Here 
		$\tan\beta = 50, \; A_0 = 0, \; m_0 = 1000 \; {\rm GeV}$}
    \label{mHp-spectrum}
  \end{figure}
Taking into account this correction yields sizable change (more than 10\%)
of predicted masses of heavy Higgs bosons and chargino (see
Fig.~\ref{mHp-spectrum}). 
This fact represents strong dependence of $\mu$ and $M_A$
on the heavy quark Yukawa couplings in certain regions of
parameter space \cite{Allanach:2003jw}.
This dependence leads to relatively large discrepancies 
\cite{Allanach:2003jw,Belanger:2005jk}
between the values of predicted masses given by different software
for calculation of MSSM mass spectrum \cite{Allanach:2001kg,MiscCodes}.
It should be noted that change due to two-loop SQCD correction to the relation
\eqref{pole2DR} exceeds these discrepancies.

	\begin{figure}
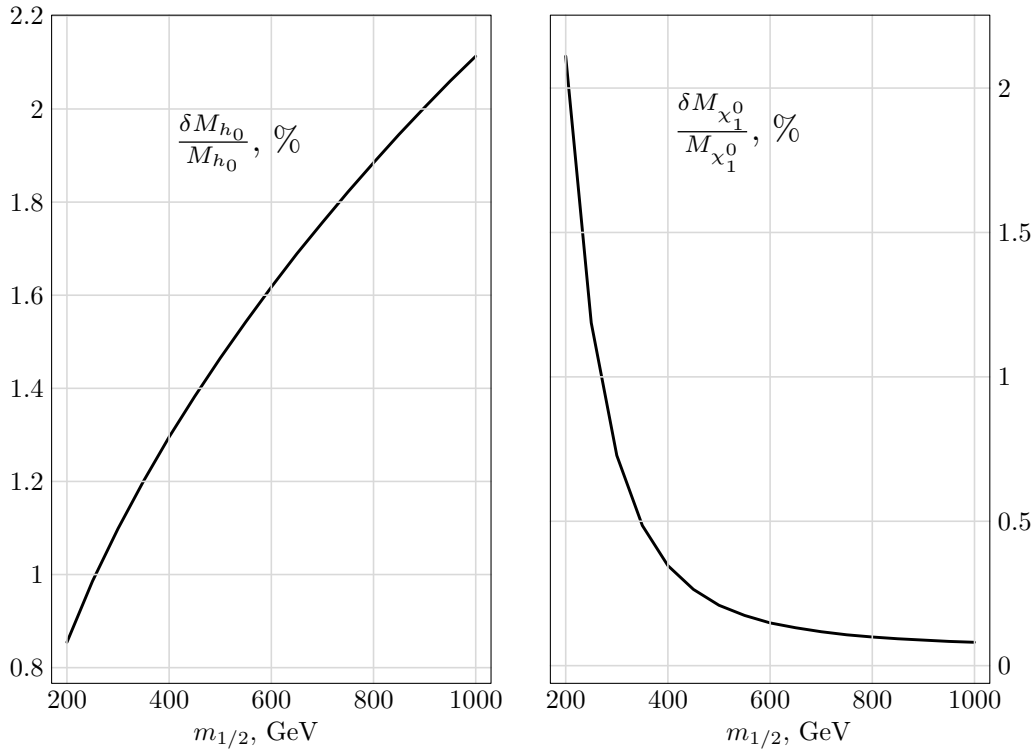

    \begin{tabular}{ll}
      \includegraphics{mh0-spectrum} & \includegraphics{mneu1-spectrum}
    \end{tabular}
    \caption{Change of predicted mass of the lightest Higgs boson and
    the lightest neutralino mass due to two-loop SQCD contribution to
    $\Delta m_t(M_Z)/m_t(M_Z)$ as a function of $m_{1/2}$.
    Here $m_0 = 1000 \, \GeV, \, \tan\beta = 50, \, A_0 = 0$.}
    \label{mh0-mneu1-spectrum}
  \end{figure}
However, masses of squarks, gluino,
and relatively light particles (lightest neutralino, lightest Higgs boson) 
do not obtain any significant changes due to above mentioned two-loop SQCD
contribution, see Fig.~\ref{mh0-mneu1-spectrum}.
\end{section}

\begin{section}{Conclusion}
\label{priplyli}
	In this paper, we presented the result of calculation of the two-loop corrections
	to the relation between pole and running masses of the $t$-quark in the
  supersymmetric QCD. We provided a numerical analysis of the
  value of these corrections in different regions of the CMSSM parameter
  space. Our analysis showed that calculated second-order terms of large mass expansion
  $\Obig{m_t^2/M_{hard}^2}$ give negligible contribution to this relation, so
  zero-order result given in \cite{Bednyakov:2002sf} provides reliable
  approximation for phenomenological studies of the MSSM. 
	Analysis given in \cite{Allanach:2003jw} demonstrates
	that there exists two regions of the MSSM 
	parameter space where accurate predictions based on
	computer codes \cite{Allanach:2001kg,MiscCodes} are difficult:
	the large {\TB} and focus-point regimes. These two regions
	require a more precise determination of the heavy quark Yukawa
	couplings, or equivalently a more precise determination of running 
	quark masses. We showed that two-loop supersymmetric QCD corrections
	give sizable contribution to \eqref{pole2DRApprox} and 
	has to be included in computer codes used to calculate MSSM spectra 
	\cite{Allanach:2001kg,MiscCodes}. 

  As a by-product of our
  calculation, we also obtained two-loop anomalous dimension of the running quark 
  $\DR$-mass in the supersymmetric QCD.
\end{section}

\begin{section}{Acknowledgements}
	The authors would like to thank M. Yu. Kalmykov for fruitful discussions
	and multiple comments. Financial support from RFBR grant \#~05-02-17603,
 grant of the Ministry of Industry, Science and Technologies of the 
 Russian Federation \#~2339.2003.02 is kindly acknowledged.
\end{section}



\begin{thebibliography}{00}
\bibitem{Bednyakov:2002sf}
A.~Bednyakov, A.~Onishchenko, V.~Velizhanin and O.~Veretin,
Eur.\ Phys.\ J.\ C {\bf 29} (2003) 87
[arXiv:hep-ph/0210258].

\bibitem{AsympExpansion}
V.~A.~Smirnov,
``Applied asymptotic expansions in momenta and masses,''
ISBN: 3540423346, Springer-Verlag (2001) (Springer tracts in modern physics, 177),
http://www.slac.stanford.edu/spires/find/hep/www?irn=4841620 \\
  A.~N.~Kuznetsov, F.~V.~Tkachov and V.~V.~Vlasov,
  arXiv:hep-th/9612037. \\
  A.~N.~Kuznetsov and F.~V.~Tkachov,
  arXiv:hep-th/9612038. \\
  F.~V.~Tkachov,
  Int.\ J.\ Mod.\ Phys.\ A {\bf 8}, 2047 (1993)
  [arXiv:hep-ph/9612284].

 \bibitem{Bednyakov:2004gr}
 A.~Bednyakov and A.~Sheplyakov,
 Phys.\ Lett.\ B {\bf 604}, 91 (2004)
 [arXiv:hep-ph/0410128].

\bibitem{Avdeev:1997sz}
L.~V.~Avdeev and M.~Y.~Kalmykov,
Nucl.\ Phys.\ B {\bf 502}, 419 (1997)
[arXiv:hep-ph/9701308].

\bibitem{Berends:1996gs}
F.~A.~Berends, A.~I.~Davydychev and V.~A.~Smirnov,
Nucl.\ Phys.\ B {\bf 478}, 59 (1996)
[arXiv:hep-ph/9602396].

\bibitem{Hahn:2001rv}
T.~Hahn and C.~Schappacher,
Comput.\ Phys.\ Commun.\  {\bf 143}, 54 (2002)
[arXiv:hep-ph/0105349].

\bibitem{prop2exp} A.~Sheplyakov,
	``prop2exp, a C++ library for asymptotic expansion of 2-loop propagator
	type integrals,''
	unpublished.
	The source code can be obtained form http://theor.jinr.ru/\~{}varg/dist

\bibitem{Bauer:2000cp}
C.~Bauer, A.~Frink and R.~Kreckel,
arXiv:cs.sc/0004015.

\bibitem{bubblesII} A.~Sheplyakov,
	``bubblesII, a C++ library for analytical and numerical evaluation of 
	2-loop vacuum integrals,''
	unpublished.
	The source code can be obtained form http://theor.jinr.ru/\~{}varg/dist

\bibitem{Davydychev:1992mt}
A.~I.~Davydychev and J.~B.~Tausk,
Nucl.\ Phys.\ B {\bf 397}, 123 (1993).

\bibitem{IntegrByParts}
  F.~V.~Tkachov,
  Phys.\ Lett.\ B {\bf 100}, 65 (1981). \\
K.~G.~Chetyrkin and F.~V.~Tkachov,
Nucl.\ Phys.\ B {\bf 192}, 159 (1981).

\bibitem{Jegerlehner:2003sp}
F.~Jegerlehner and M.~Y.~Kalmykov,
Acta Phys.\ Polon.\ B {\bf 34}, 5335 (2003)
[arXiv:hep-ph/0310361].

\bibitem{Castano:1993ri}
D.~J.~Castano, E.~J.~Piard and P.~Ramond,
Phys.\ Rev.\ D {\bf 49}, 4882 (1994) [arXiv:hep-ph/9308335].
\bibitem{Martin:1993zk}
	S.~P.~Martin and M.~T.~Vaugh:,
	Phys.\ Rev.\ D {\bf 50}, 2282 (1994)
	[arXiv:hep-ph/9311340].

\bibitem{MSSMfits}
W.~de Boer, M.~Huber, C.~Sander and D.~I.~Kazakov,
Phys.\ Lett.\ B {\bf 515}, 283 (2001). \\
H.~Baer, C.~Balazs, A.~Belyaev, T.~Krupovnickas and X.~Tata, 
JHEP {\bf 0306}, 054 (2003) [arXiv:hep-ph/0304303]. \\
W.~de Boer and C.~Sander,
Phys.\ Lett.\ B {\bf 585}, 276 (2004) [arXiv:hep-ph/0307049]. \\
J.~R.~Ellis, K.~A.~Olive, Y.~Santoso and V.~C.~Spanos,
arXiv:hep-ph/0408118.

\bibitem{Allanach:2001kg}
B.~C.~Allanach,
Comput.\ Phys.\ Commun.\  {\bf 143}, 305 (2002) [arXiv:hep-ph/0104145].

\bibitem{MiscCodes}
A.~Djouadi, J.~L.~Kneur and G.~Moultaka,
arXiv:hep-ph/0211331. \\
W.~Porod,
Comput.\ Phys.\ Commun.\  {\bf 153}, 275 (2003) [arXiv:hep-ph/0301101]. \\
F.~E.~Paige, S.~D.~Protopescu, H.~Baer and X.~Tata,
arXiv:hep-ph/0312045.

\bibitem{Allanach:2003jw}
B.~C.~Allanach, S.~Kraml and W.~Porod,
JHEP {\bf 0303}, 016 (2003)
[arXiv:hep-ph/0302102].

\bibitem{Belanger:2005jk}
  G.~Belanger, S.~Kraml and A.~Pukhov,
  arXiv:hep-ph/0502079.
\end{thebibliography}
\end{document}